\documentclass[aps,pra,reprint,groupedaddress]{revtex4-1}

\usepackage{graphicx}
\usepackage[cmex10]{amsmath}
\usepackage{amssymb}
\usepackage{algorithm}
\usepackage{algorithmic}
\usepackage{array}
\usepackage{multirow}
\usepackage{url}
\usepackage{subfig}
\usepackage{amsthm}
\newtheorem{definition}{Definition}
\newtheorem{theorem}{Theorem}
\newtheorem*{prob*}{Problem Statement}

\newcommand{\bra}[1]{\left\langle{#1}\right\vert}
\newcommand{\ket}[1]{\left\vert{#1}\right\rangle}

\begin{document}

\title{A Specification Format and a Verification Method of Fault-Tolerant Quantum Circuits}

\author{Alexandru Paler}
\email[]{alexandrupaler@gmail.com}
\affiliation{Johannes Kepler University, Linz Institute of Technology, Altenbergerstra\ss e 69, 4040 Linz, Austria}
\author{Simon J. Devitt}
\email[]{simon.devitt@uts.edu.au}
\affiliation{Centre for Quantum Software \& Information (QSI), Faculty of Engineering \& Information Technology, University of Technology Sydney, Sydney, NSW, 2007, Australia.}
\affiliation{Turing inc., Berkeley, CA, 94701 USA}

%zi undeva ca e mai bine asa deoarece am tabelul de stabilizatori si pot extrage stabilizatorii din geometrie?

\date{\today}

\begin{abstract}

Quantum computations are expressed in general as quantum circuits, which are specified by ordered lists of quantum gates. The resulting specifications are used during the optimisation and execution of the expressed computations. However, the specification format makes it difficult to verify that optimised or executed computations still conform to the initial gate list specifications: showing the computational equivalence between two quantum circuits expressed by different lists of quantum gates is exponentially complex in the worst case. In order to solve this issue, this work presents a derivation of the specification format tailored specifically for fault-tolerant quantum circuits. The circuits are considered a form consisting entirely of single qubit initialisations, CNOT gates and single qubit measurements (ICM form). This format allows, under certain assumptions, to efficiently verify optimised (or implemented) computations. Two verification methods based on checking stabiliser circuit structures are presented.

\end{abstract}

% insert suggested PACS numbers in braces on next line

\maketitle

\section{Introduction}
\label{sec:motiv}
The first generation of large scale quantum computers will have to execute fault-tolerant quantum circuits protected by quantum error correcting codes (QECC). Such circuits require large amounts of computational resources (physical qubits and execution time). The computers will be resource constrained, and the mismatch between available and required resources is one of the major hurdles that have to be overcome for quantum computations to be practical. Therefore, before being executed, fault-tolerant quantum circuits need to be optimised with regard to their computational resource overhead.

A circuit is optimised after synthesis, which means that a mathematically formulated quantum algorithm is decomposed into an equivalent quantum circuit formed of quantum gates chosen from a discrete set that can be implemented using the QECC of choice. Furthermore, synthesis refers also to the translation of non-fault-tolerant quantum (non-ft) circuits into ft-circuits. Arbitrary non-ft circuits can be transformed into fault-tolerant ones \cite{paler2017fault} of the following form: 1) single qubit initialisation; 2) CNOT gates; and 3) single qubit measurements. Such circuits are called ICM, and this work uses them as a starting point for the discussion. Without loss of generality, when referring in the following to ft-circuits their ICM form is considered.

The ICM form of the ft-circuit is an intrinsic property of the non-ft variant: arbitrary non-ft quantum gates can be decomposed into an ICM sub-circuit, and the ft-circuit is the result of ICM decomposing each non-ft circuit gate. The structure of the ft-circuit will resemble the non-ft circuit. However, in order to achieve fault-tolerance, ft-circuits have to be protected by QECCs. QECC choice is flexible, because the ft-circuit's structure is independent of the chosen QECC. Each QECC has its particularities including the realisation of fault-tolerant quantum gates: some can be easily realised (transversal construction \cite{nielsen2010quantum}), while others are more complicated (e.g. using state injection, distillation and teleportation \cite{nielsen2010quantum,BK05+}).

\subsection{Motivation}
\label{sec:motiv}

This work is motivated by a straightforward problem. Ft-circuits protected by the surface code have two forms: 1) a canonical one, which is the direct result of translating the ICM structure into surface code elements; 2) an optimised one, which is obtained by optimising the surface code protected circuit using topological compression rules \cite{fowler2012bridge,raussendorf2007topological}. 

\begin{figure*}[ht!]
\centering
\includegraphics[width=1.5\columnwidth]{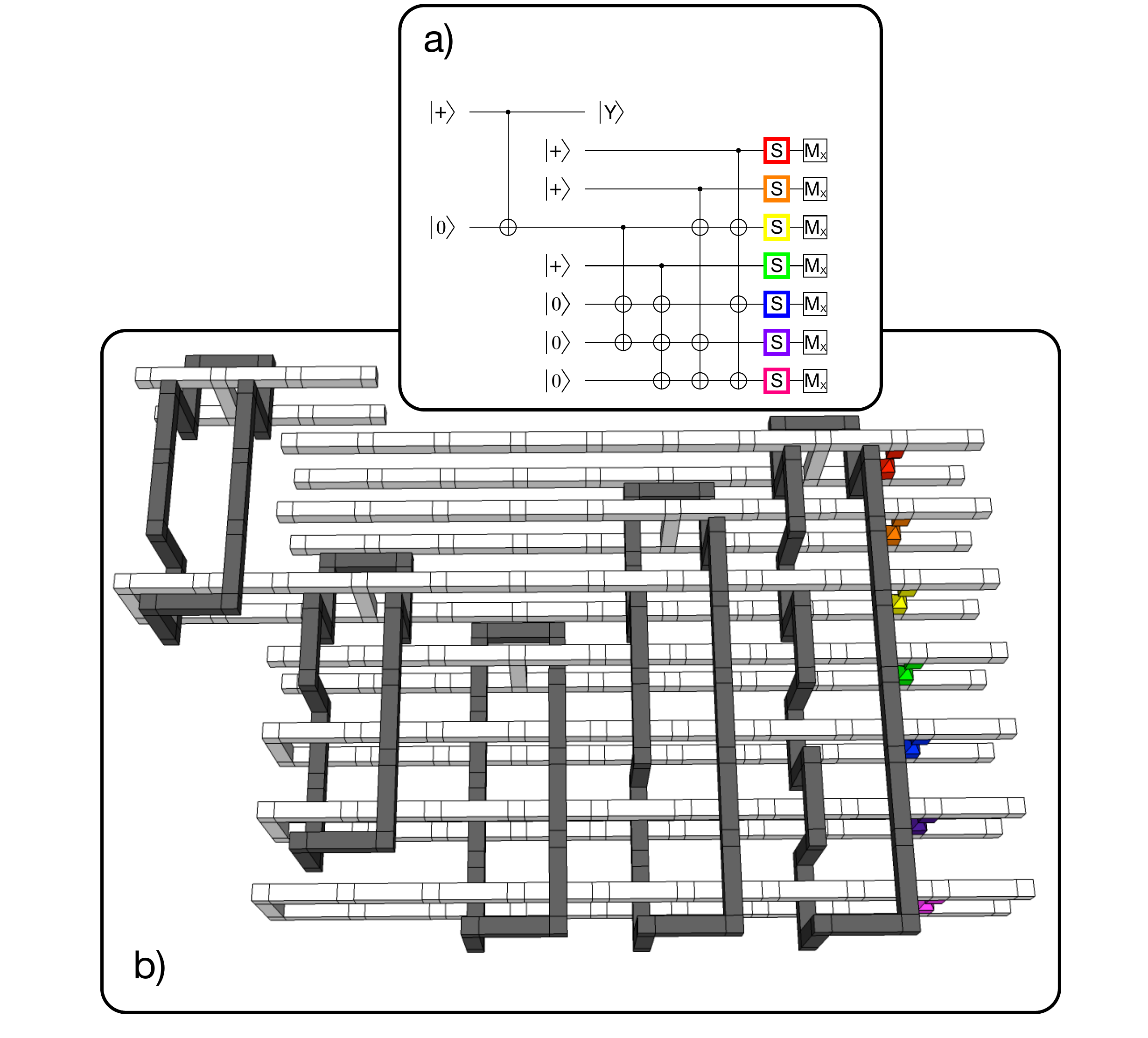}
\caption{A canonical topological circuit is constructed directly from a circuit level specification that is, in general, given in terms of the Clifford $+$ $T$ library \cite{paler2017fault}.  Our process {\em assumes} that the circuit level description has already been optimised and verified against a higher level specification and is therefore correct.  The canonical topological circuit can be verified by inspection against the circuit description but is not optimised with respect to physical qubit and time resources due to its large 3D volume.  Figure a) represents a small state-distillation circuit containing eight qubits, five multi-target CNOT gates, seven single qubits $S$-gates and seven Clifford measurements.  The output of the circuit is the single uppermost qubit.  Figure b) is the canonical topological form.  Each of the eight qubits are represented as pairs of white puzzle pieces or ``defects'' running left to right.  Each CNOT is a gray loop that braids with the relevant qubits in the circuit involved in the CNOT and the $S$-gates are the respective coloured pyramid structures \cite{paler2107fault}. }
\label{fig:canonical}
\end{figure*}

A canonical topological circuit is illustrated in Figure~\ref{fig:canonical}a) with the corresponding quantum circuit illustrated in Figure~\ref{fig:canonical}b).  This topological description of the circuit details how defects in the surface code or Raussendorf code are manipulated to enact the logic operations.  The underlying error correction processing is abstracted away in this description and is handled by separate classical components in the software stack of the quantum computer \cite{fowler2012surface, devitt2010}. It is assumed that the quantum circuit level description has already gone through a separate layer of verification procedures and is an an accurate implementation of some higher level subroutine.  The canonical form is easily verifiable against the quantum circuit itself. For the example in Figure~\ref{fig:canonical}a) each pair of white structures running left to right in the image represents the eight qubits in the original circuit, each of the five gray structures represent the five CNOT gates in the original circuit (with topological braids occurring over the relevant subset of qubits involved in each CNOT) and the seven colored pyramids represent state injection and teleportation gates for the seven $S$ gates in the original circuit.  

Once this canonical form is constructed from an input quantum circuit (the specifics of this construction can be found in Ref. \cite{paler2017fault}) it is further optimised using a series of topological compression rules \cite{fowler2012bridge,raussendorf2007topological}.  These compression rules preserve the function of the circuit (how quantum information flows and changes through the circuit) but reduce its 3D geometric volume.  The 3D volume of these structures ultimately dictate the physical resource costs (individual qubits and wall time of the quantum computer) involved in implementing them.  Consequently, the role of any optimisation technique is to shrink the physical size of the canonical quantum circuit as much as possible.  

\begin{figure*}[ht!]
\centering
\subfloat[ ]{
	\includegraphics[width=1.2\columnwidth]{ystate.pdf}
	\label{fig:ya}
}
\hspace{0.1\columnwidth}
\subfloat[ ]{
	\includegraphics[width=0.8\columnwidth]{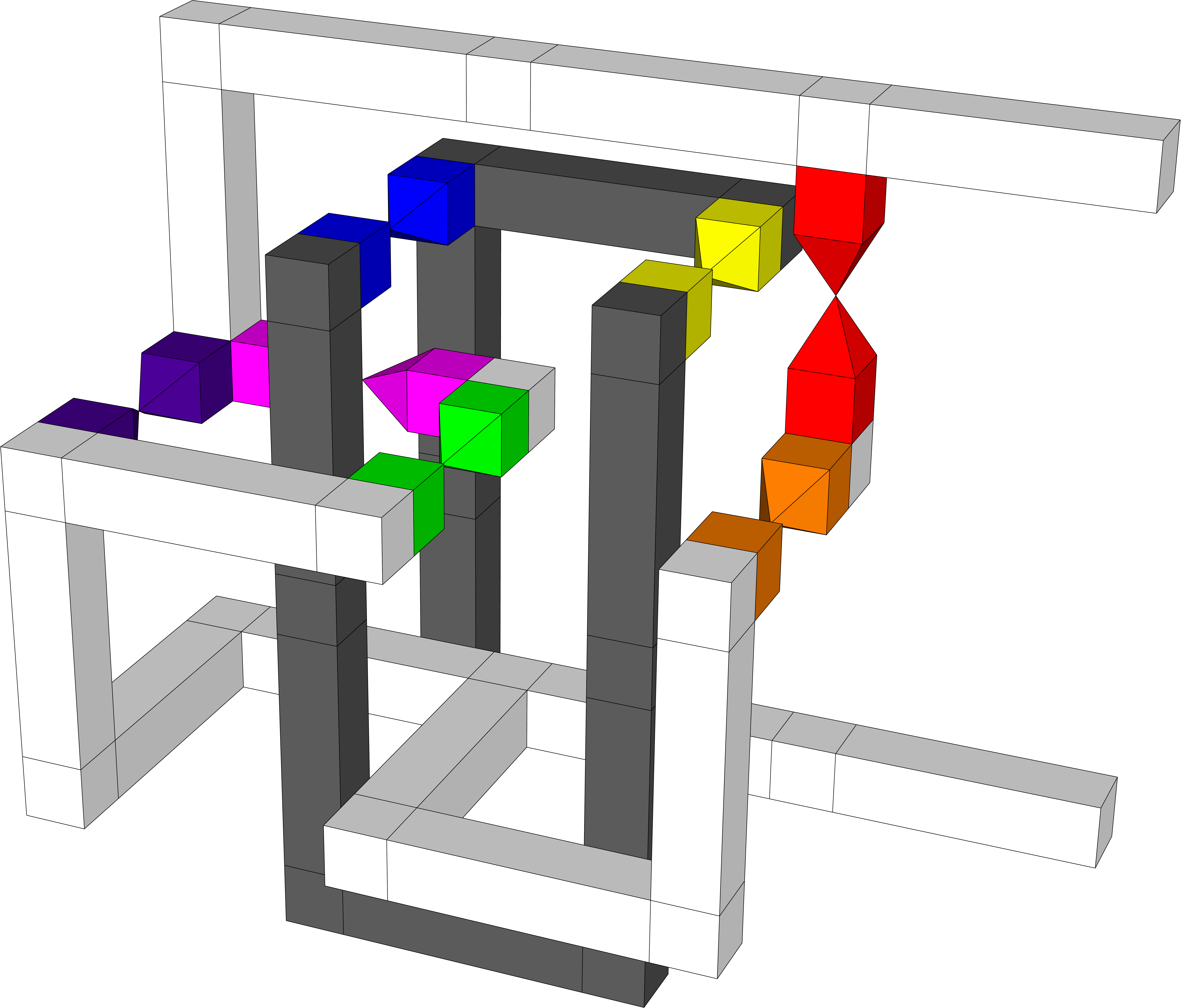}
	\label{fig:yb}
}
\caption{Representation of an ft-circuit protected by the surface code \cite{fowler2012bridge}: a) canonical; b) optimised.  Both circuits are functionally equivalent, and produce the same output, but circuit b) requires far less physical resources (physical qubits, time) than circuit a).  Circuit, a) is easily mapped to the original circuit specification, while b) is not.}
\label{fig:y}
\end{figure*}

An example of this topological compression for the circuit in Figure~\ref{fig:ya} is illustrated in Figure~\ref{fig:yb}.  This compression was performed in Ref.\cite{fowler2012bridge} and consisted of nearly 100 individual ``moves'' (see supplementary material of Ref.\cite{fowler2012bridge}).  The optimised structure of Figure~\ref{fig:yb} has a 3D geometric volume over an order of magnitude smaller than its canonical counterpart of Figure~\ref{fig:ya} while performing {\em exactly} the same computational function.  However, while Figure~\ref{fig:ya} can be compared to the original circuit specification of Figure~\ref{fig:canonical}b), the compressed version of Figure~\ref{fig:yb} cannot in an obvious way.  Given that for large quantum circuits, potentially millions of ``moves'' will be needed to transform a canonical circuit to an resource optimised form, we need to find a technique that allows us to verify the resulting topological structure against the original canonical form and hence original circuit level specification.  

The naive approach to verifying a compressed topological structure is to simply keep track of every single ``move'' that is made during compression.  After each individual ``move'', the two circuits are compared and any differences redundantly checked to ensure they satisfy valid compression rules \cite{fowler2012bridge}.  This would, computationally, be an extremely costly component of a topological quantum circuit optimizer that is already anticipated to be a complex process in the classical compilation stack for an error-corrected quantum algorithm \cite{HND17}.  Ideally we would want to take a geometric structure that is output from an automated topological circuit optimizer (which currently does not exist) and verify it without having to examine, step-by-step, the detailed record of how it was derived from the canonical structure itself.  

%The canonical forms (e.g. Figure~\ref{fig:ya}) can be easily inspected for correctness, but this is not the case for the optimised ones (e.g. Figure~\ref{fig:yb}) although both forms implement the same quantum circuit. The canonical form can be easily translated back to the original quantum circuit. However, to the best of our knowledge, it is not known how to perform such a translation for the optimised form if the compression rule sequence that was applied is not given.

Consequently, there are two identical problems: 1) verify that a compressed form is the implementation of a given quantum circuit; 2) develop a translation technique from the optimised form back to the original canonical structure. Given that option 2) is essentially the reverse of keeping a detailed record of how an optimised structure is derived, this work addresses the first problem.  By deriving a technique that verifies the optimal topological structure against the original circuit specification, we can not only apply it to the specifics of the surface and Raussendorf code but we can also generalise it further (Section~\ref{sec:verif}).

It should be noted that it is currently unknown if a compressed/optimized topological structure can be derived {\em directly} from the original circuit specification.  Current techniques start with a verified quantum circuit that has been decomposed into the Clifford $+$ $T$ gate library which is then converted to a resource inefficient canonical form \cite{paler2017fault} before undergoing compression.  

This paper therefore addresses the following problem:

\begin{prob*}
Given two ft-circuits, $ft_1$ and $ft_2$, where it is supposed that $ft_2$ is an implementation of $ft_1$, determine if $ft_1$ and $ft_2$ are functionally equivalent (is the supposition true?).
\end{prob*}

\subsection{Implementation and black box analogy}

The term \emph{implementation} from the problem statement will be explained in the following using a black box analogy. If $ft_1$ and $ft_2$ would be expressed in the quantum circuit formalism, $ft_2$ is seen as a transformation of $ft_1$ through gate identities (e.g. $HXH=Z$) which leave the number of qubits unchanged. The transformation is considered unknown, and it is not computationally feasible to backtrace all potential gate identity sequences to obtain $ft_1$. This allows to consider that, on the one hand, $ft_1$ is known (its wires and gates are visible and its functionality is captured by a functional specification) and, on the other hand, $ft_2$ seems to be hidden in a black box (it is not known how it was obtained). Thus, \emph{implementation} indicates a kind of \emph{structural relation} between $ft_1$ and $ft_2$: the latter should be obtained from $ft_1$ without changing the number of qubits. 

%rescrie
The circuit's structure is partially determined by the number of operated qubits. The black box model used in this work considers that qubit initialisations and measurements (including ancilla) are external to the box containing the circuit. The $ft_2$ circuit is applied to a known number of qubits (ancilla or not), and this indicates a compatibility between the black box model and the initial problem of determining if a canonical form circuit (an illustration of $ft_1$) is equivalent to a compressed form (an illustration of $ft_2$). Compressed circuits act like like black boxes, because no method seems to be currently known about how to \emph{read} the computation from a compressed braided circuit, and only qubit initialisations and measurements are recognisable (accessible, external to the box).

Consequently, because the number of $ft_2$ qubits can be determined, the herein presented verification \emph{assumes} that $ft_1$ and $ft_2$ operate on the same number of qubits. Otherwise, although the circuits may be functionally equivalent (e.g. $ft_1$ is the identity computation on a single qubit and uses no ancilla, but $ft_2$ includes ancillas which are used to implement the same identity computation), $ft_2$ cannot be considered an implementation of $ft_1$, and the verification will indicate non-equivalence.

The black box model allows the extension of the verification method beyond the braided formalism of surface code protected circuits. The model does not mention if error correction (or which specific code) is used within the box or not, thus the verification method is applicable as long as $ft_1$ and $ft_2$ are structurally related and have the ICM form (Section~\ref{sec:verif}). Finally, it is possible to note that the method is a \emph{weak} verification according to the classification from \cite{jordan2014strong}, because it uses information about circuit qubits.

\subsection{Related Work and Contributions}

The problem of verification is analysed in at least two contexts. First, verification is performed in the context of quantum circuit design automation: given a circuit that is known to be correct, the task is to prove that a new circuit (e.g. optimised) is equivalent to the original circuit \cite{yamashita2010fast}. Multiple approaches were investigated including verification during synthesis (compilation) \cite{amy2017verified}, SAT-based approaches \cite{yamashita2010fast} and QMDD-based approaches \cite{wille2009equivalence}. Second, verification is concerned with checking whether quantum computers are indeed producing correct results \cite{gheorghiu2017verification}, where a verifier with (almost) classical computing capabilities tries to determine if a black boxed machine (a prover which is an untrusted entity \cite{aharonov2013quantum}) is falsifying the results of a quantum computation. The methods developed in this context are used to test whether the quantum computer is quantum and even whether it computes correctly \cite{barz2013experimental}. More recently, some methods for verifying the untrusted quantum computer have considered the use of stabiliser computations (e.g. \cite{hayashi2015verifiable,fujii2017verifiable}). The major distinction between the two interpretations of verification is that the first assumes trusted quantum hardware, whereas the second does not.

This work is based on the design automation interpretation of the verification problem.
 However, the methodology presented in this work is somehow related to \cite{fujii2017verifiable}, because it uses stabilisers for verifying optimised ft-circuits.

Previous results regarding the problem stated in this work were presented in \cite{paler2014cross}. In that work ft-circuits are protected by topological QECCs (e.g. surface codes), and the validity of optimised (compressed) ft-circuits is based on mapping the circuits to an underlying lattice of physical qubits (necessary for the QECC) and simulating the resulting lattice circuit. The ft-circuit was considered high-level and the lattice low-level. It was recognised that checking the validity of the high-level optimised circuit can be performed efficiently by simulating only low-level stabiliser circuits.

This work advances the results from \cite{paler2014cross}. First, the specification format was not defined clearly, and this work will fill this gap. Second, the ft-circuits were not considered in their ICM form, and this affected the generality of the approach, because it did not take into consideration that some ft-circuit gates require probabilistic corrections which cannot be tracked in the Pauli frame \cite{paler2014software} (e.g. see Section~\ref{sec:measrules}). Third, we extend the results from \cite{paler2017fault} and show that there are two distinct types of ICM forms (see Section~\ref{sec:rotated}). Fourth, the methods presented herein do not require any lower level circuit simulations. Finally, we argue that the specification format and the verification method can be used to replace (to a certain degree) tomographic methods for testing quantum computations implemented in hardware.

\section{Methods}

This work proposes a quantum circuit specification format which includes all necessary information to represent ft-circuits. This section presents a step-by-step derivation of the proposed specification format. The important characteristics of a specified ft-circuit are determined starting from a classification of ancilla qubit types. Afterwards, the concept of \emph{stabiliser truth table} is introduced. Finally, in conjuction with the ICM form of ft-circuits, the effect of single qubit measurements is briefly analysed. This allows to clearly define the specification format, and to highlight one of its applications, which is verification.

\subsection{Classification of Ancilla Qubit Types}
\label{sec:deriv}

Quantum information processing can be analysed using a black-box analogy of quantum circuits. A black-box circuit operates on \emph{IO-qubits} and various \emph{ancillae} types: IO-qubits are the interface to the box, and ancillae are internal to the box. The main difference between the IO- and the ancilla qubits is that, the initialisation and measurement of the first is flexible (multiple bases are allowed), whereas ancillae are initialised and measured into specific bases.

Both ft- and non-ft circuits employ ancillae. In a non-ft circuit, \emph{computational ancillae} are temporary workbenches (e.g. in quantum arithmetic circuits). Such ancillae are, by definition, initialised in a determined state, used during the computation, their state is reversed to the initial one, and finally the qubits are implicitly measured.

\begin{figure}[t!]
\centering
\includegraphics[width=\columnwidth]{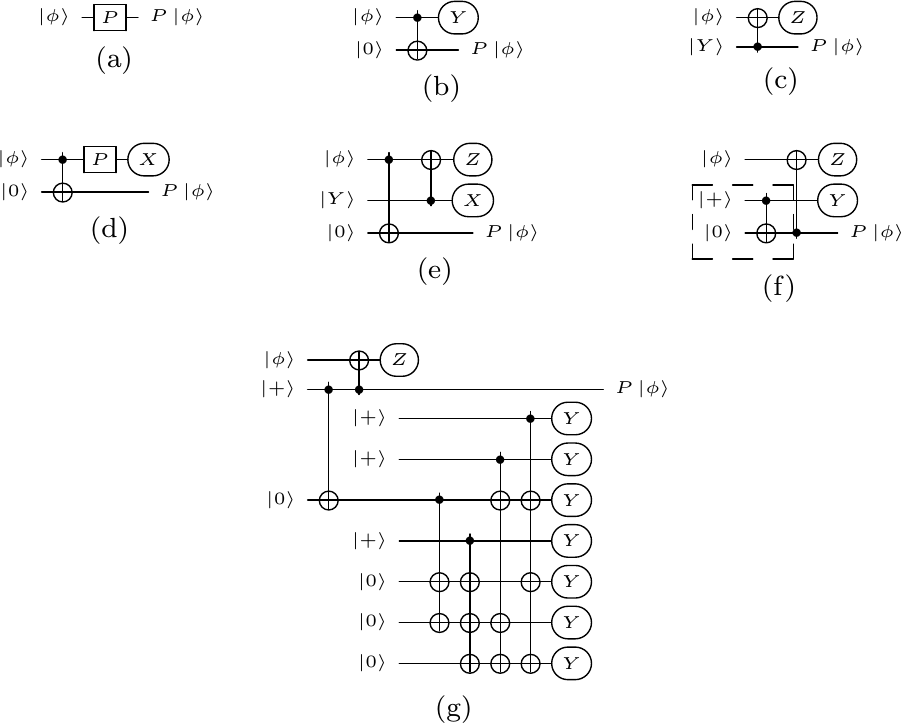}
\caption{\label{fig:2}\small $P$-gate implementations: b) rotated-measurement (measure in Y basis); c) rotated-initialisation (initialise in $Y$ basis). Transformations between implementations: d) rewrite circuit \ref{fig:2}b as \ref{fig:2}d; e) replace the $P$ gate with circuit \ref{fig:2}c; f) circuit \ref{fig:2}c is transformed into \ref{fig:2}f. Increase measurement fidelity: g) circuit \ref{fig:2}c using a distillation procedure.}
\label{fig:2}
\end{figure}

In particular, ft-circuits are commonly analysed from the perspective of the Clifford $+$ $T$ gate set. As a result, two additional ancillae types exist: \emph{teleportation} and \emph{distillation}. Quantum information teleportation is a computational primitive used in the non-transversal ft-gate constructions of the $T$, $P$ and $V$ rotation gates. Gate functionalities are implemented by entangling an \emph{initial qubit} (IO-qubit or ancilla) with a \emph{teleportation ancilla}. Distillation ancillae are used during the preparation of rotated basis initialisation or measurement.

\begin{figure}
\centering
\includegraphics[width=0.8\columnwidth]{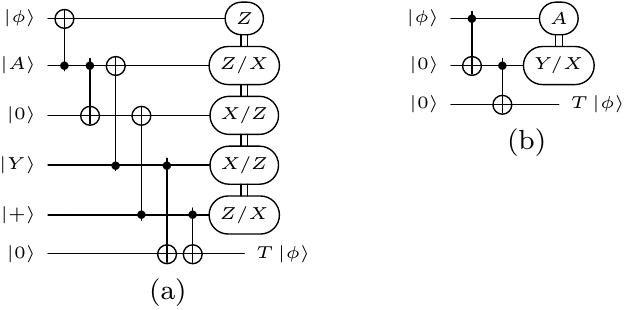}
\caption{\small Fault-tolerant $T$ gate implementation. The measurement result of the top qubit determines the measurement basis on the other qubits: a) rotated-initialisation \cite{fowler2012time}; b) rotated-measurement.}
\label{fig:corr}
\end{figure}

\subsubsection{Rotated Initialisation and Rotated Measurement}
\label{sec:rotated}

\begin{figure}
\centering
\includegraphics[width=0.7\columnwidth]{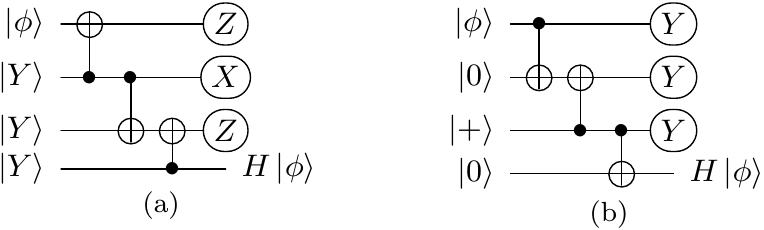}
\caption{Circuits for performing a Hadamard gate using teleportation ancillae, CNOT gates, and a measurements. a) rotated initialisation of ancillae; b) rotated measurement of ancillae.}
\label{fig:options}
\end{figure}

Due to the manner in which ft-gates are implemented through teleportation, there are two techniques for implementing them: 1) \emph{rotated-initialisation} (Figure~\ref{fig:options}a), where the teleportation ancilla is initialised into a rotated basis ($A$ or $Y$, Equations~\ref{eq:a} and \ref{eq:y}) and the initial qubit is measured in either the $X$ or $Z$ basis; or 2) \emph{rotated-measurement} (Figure~\ref{fig:options}b), having the teleportation ancilla initialised into either $X$ or $Z$ and the initial qubit measured in $A$ or $Y$.

\begin{align}
\ket{A_0}&=\frac{1}{\sqrt{2}}(\ket{0} + e^{i\frac{\pi}{4}}\ket{1});\ket{A_1}=\frac{1}{\sqrt{2}}(\ket{0} - e^{i\frac{\pi}{4}}\ket{1})\nonumber\\
A &= \ket{A_0}\bra{A_0} - \ket{A_1}\bra{A_1}\label{eq:a}\\
\ket{Y_0}&=\frac{1}{\sqrt{2}}(\ket{0} + i\ket{1});\ket{Y_1}=\frac{1}{\sqrt{2}}(\ket{0} - i\ket{1})\nonumber\\
Y&=\ket{Y_0}\bra{Y_0} - \ket{Y_1}\bra{Y_1}\label{eq:y}
\end{align}

An ft-circuit can be transformed with constant overhead (a supplemental ancilla and a single CNOT) from one technique to the other (e.g. Figure~\ref{fig:2}d where the $P$ gate is replaced by the circuit from Figure~\ref{fig:2}c) \cite{HND17}. The transformation is based on the observation that a rotated basis measurement is equivalent to first applying rotation gates before measuring in the $X$ or $Z$ basis.

Consequently, ft-gate construction techniques are equivalent. It is possible to rewrite an entire rotated-initialisation circuit into a rotated-measurement one, or viceversa, by: 1) reversing CNOT directions, and 2) switching the interpretation of initialisation and measurement (see the differences between Figure~\ref{fig:2}b and \ref{fig:2}c).

High fidelity measurements or initialisations, necessary for reaching the fault-tolerant threshold, are achieved through distillation procedures (applicable to both $A$ and $Y$ states). The procedures, magic state distillation protocols \cite{bravyi2005universal}, are implemented as stabiliser sub-circuits using \emph{distillation ancillae}. These ancillae are initialised into the $X$ or $Z$ basis, are interacted during the protocol, and measured in a rotated basis (e.g. Figure~\ref{fig:2}g). A distillation sub-circuit outputs (probabilistically) a higher fidelity rotated state required as input (teleportation ancilla) to a rotated-initialisation gate (e.g. a $P$-gate in Figure~\ref{fig:2}g).

\subsubsection{Classification Summary}

IO-qubit and ancillae properties can be summarised: IO-qubits are initialised into $X$ or $Z$, and are measured into $X$ or $Z$. Computational ancillae are implicitly measured in the initialisation basis (e.g. if initialised into $\ket{0}$, the measurement is in the $Z$ basis). Teleportation and distillation ancillae are measured in a basis different to initialisation: initialised into a rotated basis and measured in $X$ or $Z$, or vice-versa.

There is a similarity between all ancillae types: they are interacted entirely through CNOT gates. Additionally, on the one hand, the IO-qubits, the computational and the teleportation ancillae are a priori known elements of the ft-circuits. On the other hand, distillation ancillae are dynamically included into circuits whenever the initialisation/measurement fidelities are low. The existence of distillation ancillae is a posteriori established: these are not required if state fidelities are sufficiently high.

\subsection{Stabiliser Truth Table}
\label{sec:stt}

Ft-circuits consist of a stabiliser and a non-stabiliser partition. The contents of each each partition depends on the chosen ft-gate implementation:
\begin{itemize}
\item rot.-meas. - stabiliser: initialisations and CNOTs; non-stabiliser: measurements;
\item rot.-init. - stabiliser: CNOTs and measurements; non-stabiliser: initialisations.
\end{itemize}

The behaviour of the stabiliser partition is captured by a so-called \emph{stabiliser truth table}, which can be constructed by conjugating input stabilizers with the unitary representing an all Clifford circuit. The table's structure is introduced by the example of the smallest possible ft-circuit: a CNOT gate. This circuit contains two IO-qubits and no ancillae. Its inputs are initialised in either the $X$ or the $Z$ basis, and its outputs are measured in the same two bases. Table~\ref{tbl:1} is the stabiliser truth table ($c$ indicates the control and the $t$ the target qubits) representing the following stabiliser transformations:
\begin{align*}
X_cI_t \rightarrow X_cX_t; I_cX_t \rightarrow I_cX_t;\\
Z_cI_t \rightarrow Z_cI_t; I_cZ_t \rightarrow Z_cZ_t;\\
\end{align*}%

\begin{definition}
A stabiliser truth table is the description of the functionality implemented by an ft-circuit stabiliser partition. The table enumerates as rows each \emph{relevant} circuit input and its corresponding output.
\label{def:2}
\end{definition}

\begin{table}[t!]
\centering\small
\begin{tabular}{r|cccc}
Nr. & $i_c$ & $i_t$ & $o_c$ & $o_t$\\
\hline
1 & X & I & X & X\\
2 & I & X & I & X\\
3 & Z & I & Z & I\\
4 & I & Z & Z & Z
\end{tabular}
\caption{\label{tbl:1}The stabiliser truth table of a CNOT.}
\end{table}

Stabiliser truth tables are obtained by conjugating the circuit IO-qubit stabilisers (X or Z) through the gates from the CNOT region. As a result, a truth table consists of two regions (input stabilisers and output stabilisers) corresponding to the CNOT region of the ft-circuit. Definition~\ref{def:2} shows the similarity between a stabiliser and a Boolean truth table, but in contrast to the latter, the stabiliser version has a major advantage: it has a linear length in the number of qubits considered (see discussion in the section detailing its construction). The stabiliser truth table includes all qubit types except the distillation ancillae, because: a) their existence is determined only during the computation; b) distillation sub-circuits do not influence the computation.

\subsubsection{Stabiliser Truth Table Operations}

The usual stabiliser notation can be forced and stabiliser truth table rows can be handled as usual stabilisers. All possible circuit states can be inferred/computed using stabiliser table row additions and multiplications. For example, the expected output after inputing $X_cX_t$ to a CNOT results by multiplying the corresponding rows from Table~\ref{tbl:1} (superscript $i$ and $o$ indicate inputs and outputs):
\begin{align}
S_1=(X_c^iI_t^iX_c^oX_t^o)(I_c^iX_t^iI_c^oX_t^o) = X^i_cX^i_tX^o_cI^o_t\label{eq:mult}
\end{align}

The following sections will show that, similar to measurement-based computing \cite{RBB03} where qubit initialisation basis and qubit measurement order determine the implemented computation, the output state of an ft-circuit is determined by the sequence of table rows operations. For the beginning, the $T$- ft-gate is a straightforward example. The non-ft $T$ gate would transform $X \rightarrow (X+Y)/\sqrt{2}$ and $Z \rightarrow Z$, but the ft-gate implementation will be structurally equivalent to the circuit from Figure~\ref{fig:2} (has the $Y$ basis replaced by an $A$ basis). Therefore, in this case, stabiliser transformations are partially determined by the CNOT region of the circuit.

The expression $S_1$ shows that it is possible to express the transformation of the IO-qubit input stabiliser $X$ (upper qubit, before the CNOT target) to an output stabiliser $X$ (bottom qubit, after the CNOT's control). Simultaneously, the stabiliser superposition resulting after the ft-gate application is expressed by $S_3$ and $S_1$ (row addition shows that $X_i^t$ can result to $Y^o_cI^o_t$ and $X^o_cI^o_t$).

\begin{align*}
S_2 &= (X_c^iI_t^iX_c^oX_t^o)(Z_c^iI_c^iZ_c^oI_t^o) = Y^i_cI^i_tY^o_cX^o_t\\
S_3 &=S_1(I_c^iX_t^iI_c^oX_t^o) = Y^i_cX^i_tY^o_cI^o_t\\
(S_3 + S_1)/\sqrt{2} &= ((Y^i_cX^i_tY^o_cI^o_t) + (X_c^iX_t^iX_c^oI_t^o))/\sqrt{2}
\end{align*}

\subsubsection{Efficient Construction}
\label{sec:sttconst}

The stabiliser truth table can be efficiently determined. The table is the result of simulating the stabiliser partition using all the basis input states. The simulation is straightforward for the rotated measurement construction: the non-stabiliser bases are at the end of the circuit, so simulation will stop right before measurements.

For rotated initialisation the construction seems more difficult. Not all the inputs are stabiliser states. Although the $A$ and the $Y$ matrices are members of the Clifford group \cite{bravyi2005universal}, the eigenvectors of $A$ represent non-stabiliser states. Therefore, in this ft-gate construction, the teleportation ancillae will be stabilised by superpositions (e.g. $\ket{A_0}$ is stabilised by $(X+Y)/\sqrt{2}$). Computing the truth table, having non-stabiliser inputs, results in an exponential overhead which is difficult to mitigate \cite{garcia2013quipu}. 

However, it is possible to efficiently construct a stabiliser truth table by temporarily replacing the rotated teleportation ancillae initialisations with $X$ and $Z$ basis initialisations. We motivate this decision by two equivalent arguments based on the observation that the truth table expresses ft-circuit basis states. 

The first argument: due to $(X+Y)/\sqrt{2}=(X+iXZ)/\sqrt{2}$, it is sufficient to compute the $X$ and $Z$ stabiliser transformations originating \emph{after} a teleportation ancillae initialised into the $A$ basis: the stabiliser superposition can be computed from the two $X$ and $Z$ transformations (similarly to Equation~\ref{eq:mult}). The computation would still introduce an exponential overhead, but the computation itself is not required, because it is a direct result of qubit initialisations and measurements (see the section detailing measurement rules and specification definition).

The second argument: the circuit in Figure~\ref{fig:2}f (analogous to the circuit in Figure~\ref{fig:2}c) constructs a Bell state on the second and third qubits (stabilised by $XX$ and $ZZ$, see dotted box in the figure). If rotated-initialisation circuits would be rewritten as rotated-measurements, the two resulting teleportation ancilla \emph{introduce two rows into the truth table}: an $X$ and a $Z$ stabiliser transformation.

Therefore, in a $q$-qubit ft-circuit, each IO-qubit is initialised in either $I$ (the identity matrix), $X$ or $Z$, such that there are $3^q-1$ possible input states ($I_1\ldots I_q$ needs not to be included). However, at most $2q$ basis states are relevant. Each IO-qubit needs to be sequentially initialised into $X$ and $Z$, while keeping the remaining IO-qubits initialised into $I$. This consideration introduces two rows into the table for each IO-qubit. Ancilla qubits have a predetermined initialisation basis: a) rotated initialisation ancillae are simulated like IO-qubits (sequential $X$ and $Z$ initialisation) and two table rows are introduced into the table; b) rotated measurement ancillae introduce a single table row, because they are initialised in \emph{either} the $X$ or $Z$ basis.

\subsection{Measurement Rules}
\label{sec:measrules}

Finally, the measurement of qubits in fault-tolerant quantum circuits is analysed before introducing the specification format.

Fault-tolerant rotation gate implementations are probabilistic because they use teleportation mechanisms. Thus, gate outputs require a correction indicated by the measurement result of the initial qubit. An incorrect application of $P$ or $V$ is corrected a posteriori by the application of Pauli gates \cite{paler2014software}, but this does not hold for incorrect $T$ gate applications. Such gates need $P$ gate corrections \cite{fowler2012time} (Figure~\ref{fig:corr}).

The measurement basis of teleportation ancillae is classically controlled through a function of the initial qubit measurement. In Figure~\ref{fig:corr}b, if the $A$ basis measurement results in eigenvalue -1 it is an indication that $T^\dagger$ was applied instead of $T$, and the second measurement will be performed in the $Y$ basis (a correctional $P$ gate is implemented). Otherwise, eigenvalue 1 indicates the correct application of $T$, the second measurement is in the $X$ basis, thus only teleporting the intermediary state from the middle ancilla to the circuit's output.

%Measurement rules can be expressed as tuples of the form $(q_1, B_1, q_2, B_2, B_3)$ interpreted as ``if the result after measuring qubit $q_1$ in basis $B_1$ has eigenvalue $1$, then qubit $q_2$ is measured in basis $B_2$, otherwise in basis $B_3$''. In Figure~\ref{fig:corr}b, the tuple for the A measurement is $(q_1, A, q_2, X, Y)$. A tuple of the form $(q_1, B_1, \emptyset, \emptyset, \emptyset)$ represents measurements that do not generate classical control signals. It can be also seen, that a set of measurement tuples specifies a \emph{measurement sequence}: some qubits can be measured only after other qubits were already measured.

\section{Results}

The previous analysis offers the necessary elements for defining a specification format. Quantum circuits are generally specified as gate lists. In contrast, classical circuits are specified in various manners including Boolean truth tables and gate lists. Generating classical circuit truth tables is an exponentially complex task, which is even more complicated for quantum circuits due to the non-discrete nature of qubit states. 

\subsection{The Specification}
\label{sec:spec}

In the following the stabiliser truth table is a central component of the specification.

\begin{definition}
A quantum circuit specification is the tuple $(ST,I,O)$, where $ST$ is a stabiliser truth table, $I$ is the set of qubit initialisation basis, and $O$ is a \emph{list} of qubit measurement basis rules.
\label{def:1}
\end{definition}

Definition~\ref{def:1} illustrates the relation between teleportation-based, measurement-based and fault-tolerant quantum computing \cite{childs2005unified}. Both sets, I and O refer only to ancillae (computational or teleportation), because IO-qubit measurement is flexible. The rules defined in $O$ express the dynamics introduced by the probabilistic nature of measurements: the measurement basis of a qubit depends on previous qubit measurement results.

Due to the structure of the tuples, $O$ specifies a \emph{measurement order} of tuples of the form $(q_1, B_1, q_2, B_2, B_3)$ interpreted as ``if the result after measuring qubit $q_1$ in basis $B_1$ has eigenvalue $1$, then qubit $q_2$ is measured in basis $B_2$, otherwise in basis $B_3$''. The tuple $(q_1, B_1, \emptyset, \emptyset, \emptyset)$ represents measurements that do not generate classical control signals.

For example, the specification of an uncorrected T ft-gate contains the initialisation set $I_t=\{(q_2, A)\}$ and the measurement set $O_t=\{(q_1,Z,\emptyset,\emptyset,\emptyset)\}$. The stabiliser truth table would be again Table~\ref{tbl:1}.

The output state could be constructed in a \emph{systematic and determined} way by using information about teleportation ancillae initialisation basis: for Y, \emph{row multiplications} are required, and for A, \emph{row multiplications and additions}. However, the output state is not required. Skipping its computation avoids an exponential representational and computational overhead of the circuit quantum state space. The functionality of the circuit is determined in a computationally functional manner by the measurement order from $O$. That set is an expression of the initial non-ft-circuit gate list. 

\subsection{Verification}
\label{sec:verif}

The specification format can be used to solve the problem stated in Section~\ref{sec:motiv}.

\begin{theorem}
An ft-circuit implementation passes the verification against a specification $Spec=(I,ST,O)$, if and only if: 1) all of its ancillae are initialised according to $I$, 2) its stabiliser partition supports $ST$, and 3) all its ancillae are measured according to $O$.
\label{def:3}
\end{theorem}

A circuit which satisfies the three properties of $Spec$ is an implementation of the specification. Furthermore, the verification, as mentioned in Section~\ref{sec:motiv}, assumes that the ft-circuit and the specification ($Spec$) refer to the same number of qubits. Under this assumption, the opposite direction of the theorem requires only the proof of the second criterion. This can be shown using a method similar to reversible circuit equivalence checking \cite{yamashita2010fast,wille2009equivalence} (following section).

\subsection{Ft-circuits verification: Equivalence Checking}

Definition~\ref{def:1} was based on the observation that ft-circuits consist entirely of CNOT gates and, as a result, the task of \emph{ft-circuit optimisation} is to minimise the number of such gates. Thus, as long as the circuits have the same number of qubits initialised/measured in the same basis, ft-circuit equivalence verification is efficient, because the second criterion in Theorem~\ref{def:3} is shown through \textbf{\emph{stabiliser circuit equivalence}: checking the stabilisers from the truth table} (as defined in Section~\ref{sec:spec}).

The following two examples illustrate the equivalence checking technique: the first uses the quantum circuit formalism, and the second is an application on compacted braided structures.

\subsubsection{Circuit level verification}

\begin{figure}
\centering
\includegraphics[width=0.7\columnwidth]{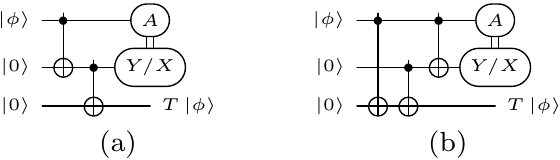}
\caption{Circuit level verification example for the ICM T gate: a) original circuit used to derive the specification; b) transformed circuit assumed to be existing in a black box.}
\label{fig:tverif}
\end{figure}

The rotated basis measurement ICM form of the T gate implementation (Figure~\ref{fig:tverif}a) can be rewritten by commuting the CNOT gates into the circuit from Figure~\ref{fig:tverif}b). According to the problem statement, Figure~\ref{fig:tverif}a is a potential illustration of $ft_1$ and Figure~\ref{fig:tverif}b of $ft_2$ existing in a black box with visible inputs and outputs. The specification $Spec=(I, ST, O)$ will be derived from $ft_1$:
\begin{align*}
I &= \{(q_2, Z), (q_3, Z)\}\\
O &= \{(q_2, A, q_3, X, Y)\}\\
ST &=
\begin{tabular}{r|ccc|ccc}
\multirow{2}{*}{Nr.} & \multicolumn{3}{|c}{Inputs} & \multicolumn{3}{|c}{Outputs}\\
 & q1 & q2 & q3 & q1 & q2 & q3\\
\hline
% * <sd@turingquantum.com> 2018-07-08T01:47:06.767Z:
%
% ^.
1 & X & I & I & X & X & X\\
2 & Z & I & I & Z & I & I\\
3 & I & Z & I & Z & Z & I\\
4 & I & I & Z & I & Z & Z
\end{tabular}
\end{align*}

Although, for the purpose of this example, the gate sequence of $ft_2$ is known, it is assumed that only its inputs and outputs can be accessed. It is necessary to check (verify) that the stabiliser transformations from the $Spec$ are supported by $ft_2$.

\subsubsection{Verification of compressed braided stucture}

Simultaneously, the verification is efficient also, because ft-circuit optimisation does not change the measurement sequence determined by $O$. Such a change is analogous to executing a different non-ft circuit gate list. Additionally, it would not be of benefit to modify the initialisation and measurement sets without reducing the number of circuit qubits. The number of io-qubits is fixed through the algorithm specification (e.g. a quantum adder on 4 qubits) and only the ancillae number could be optimised. This is a task of \emph{non-ft circuit optimisation} and there are two possible strategies. Firstly, there are various investigations about the trade-off between a non-ft circuit's number of gates and the number of computational ancillae \cite{miller2010reducing}. Secondly, the number of teleportation ancillae is a function of the number of single qubit rotational gates. Minimising the number of such gates is performed through optimal Clifford $+$ $T$ circuit synthesis or efficient arbitrary gate decompositions \cite{ross2014optimal, kliuchnikov2013fast}. Reducing the number of distillation ancillae would also be the entirely separate problem of \emph{optimal distillation circuits} which is a subproblem of non-ft circuit optimisation.

\subsection{Implementation Verification}

A circuit designed and optimised for a specific quantum computing architecture will be implemented and executed in hardware. 

The verification of an implemented quantum circuit is generally performed through tomographic methods which are computationally expensive and not scalable for large circuits. The practicality of a quantum circuit specification is directly related to the complexity of verifying if implemented circuits are conforming to their specification. We argue that implemented \textbf{circuits could be efficiently verified in a setting where}: 1) the quantum computer architecture allows flexible qubit initialisations or measurements, thus supporting also rotated basis such as A or Y, and  2) the \textbf{initialisations and the measurements are trusted}.

Assuming that the inputs and the outputs of the implementation are configurable, it is possible to efficiently verify the stabiliser truth table of the implementation using the same arguments as in Section~\ref{sec:stt}. Verification is probabilistic, as each truth table row needs to be verified multiple times, but there is a constant number of repetitions required to polynomially increase the overall verification probability. This happens because the main issue with implementation verification is related to initialisations and the measurements.

The difficulty of verification is transformed into the problem of trusting the inputs and outputs of the circuits. From an engineering point of view, the devices used to initialise or to measure qubits are identical components of the computer. Ensuring their correct functionality is a general, and not a circuit, implementation problem. Therefore, trusting quantum IO-devices is similar to how one trusts the current transistor technology: once the technology is mature, individual IO-device instance will need testing, but this is performed separately from the quantum computer. Furthermore, we did not assume that any component of the quantum computer will have a Byzantine behaviour. The proposed implementation verification method assumes that hardware can be trusted, which is entirely opposite to approaches like \cite{gheorghiu2017verification}.

\section{Conclusion}

Ft-circuits are necessary during the implementation of practical large-scale quantum computations. Their regular structure is the foundation of a specification format consisting of a stabiliser truth table and two sets regarding qubit initialisation and measurement. The major advantage of the specification is that it shows that ft-circuits can be efficiently verified using a conceptually simple method. Future work will result in the development of a scalable verification software for large scale ft-circuits.

\begin{acknowledgments}
A.P. acknowledges support from the Linz Institute of Technology (Project CHARON). 
\end{acknowledgments}

\bibliographystyle{plain}
\bibliography{2page}

\end{document}